\documentclass[12pt]{article}
\usepackage{eurosym,cite}
\usepackage[left=0.7in, right=0.7in, top=0.7in, bottom=0.7in]{geometry}
\usepackage{hyperref}
\usepackage{mathtools}
\usepackage{mathrsfs}
\usepackage[utf8]{inputenc}
\usepackage[T1]{fontenc}
\usepackage{mathrsfs}
\usepackage{amssymb}
\usepackage{tikz}
\usepackage{tikz-cd}
\usepackage{pgfplots}
\usepackage{color}
\usepackage{subfigure}
\usepackage{graphicx}
\usepackage{dcolumn}
\usepackage{bm}
\usepackage{orcidlink}
\usepackage{pstricks}

\usepackage{pgfplots}
\pgfplotsset{compat=1.16}
\usepackage{siunitx}
\usepackage{tikz}
\usepackage{tikz-cd}
\usepackage{pgfplots}
\usepackage{overpic}

\usepackage[absolute,overlay]{textpos} 

\begin{document}
	
		\begin{center}
		\textbf{\LARGE Strengthening the Uncertainty and the Reverse Uncertainty Relation Limits \\}
		\bigskip

		M.Y.Abd-Rabbou \orcidlink{0000-0003-3197-4724} $^{a,b}$ \textit{\footnote{e-mail:m.elmalky@azhar.edu.eg}},  Cong‑Feng Qiao \orcidlink{https://orcid.org/0000-0002-9174-7307} $^{a,c}$ \textit{\footnote{e-mail:qiaocf@ucas.ac.cn}}
		
		{\footnotesize  
			$^{a}${School of Physics, University of Chinese Academy of Science, Yuquan Road 19A, Beijing, 100049, China}\\
			$^{b}${\footnotesize Department, Faculty of Science, Al-Azhar University, Nasr City 11884, Cairo, Egypt.}\\
			$^{c}$ {\footnotesize Key Laboratory of Vacuum Physics, University of Chinese Academy of Sciences, Beijing 100049, China}}
		
	\end{center}

\date{\today}
\def\be{\begin{equation}}
  \def\ee{\end{equation}}
\def\bea{\begin{eqnarray}}
\def\eea{\end{eqnarray}}
\def\f{\frac}
\def\n{\nonumber}
\def\l{\label}
\def\p{\phi}
\def\o{\over}
\def\R{\rho}
\def\pa{\partial}
\def\om{\omega}
\def\na{\nabla}
\def\P{$\Phi$}

\begin{abstract}
	Uncertainty relations are pivotal in delineating the limits of simultaneous measurements for observables. In this paper, we \textcolor{black}{derive four novel uncertainty and reverse uncertainty relations} for the sum of variances of two incompatible observables, leveraging the mathematical framework of the Maligranda inequality. These relations are shown to \textcolor{black}{provide tighter bounds than several well-known existing relations}. Furthermore, we extend these results to multi-observable scenarios by employing an inequality from M. Kato et al., deriving generalized uncertainty relations that similarly exhibit enhanced precision. The incorporation of the phase angle of the measurement state contributes to strengthening the derived inequalities. Comparative analyses with prior studies confirm the \textcolor{black}{effectiveness of our inequalities in two-observable systems} via three illustrative examples.
\end{abstract}

\section{Introduction}
The investigation of uncertainty relations in quantum mechanics plays a crucial role in understanding the fundamental limits of measurement and the nature of quantum states \cite{wheeler2014quantum,erhart2012experimental}. The Heisenberg uncertainty principle establishes that specific pairs of observables cannot be simultaneously measured with arbitrary precision \cite{heisenberg1927anschaulichen}. This intrinsic limitation has led to the development of various formulations and generalizations, particularly in the context of stronger and reverse uncertainty relations \cite{PhysRevLett.111.230401,PhysRevA.91.042133,song2016stronger,RevModPhys.89.015002,PhysRevA.95.022112,PhysRevA.100.022116,liu2024quantum}.  For two such observables $\hat{A}$ and $\hat{B}$, the Heisenberg–Robertson uncertainty relation asserts that \cite{PhysRev.34.163} 
\begin{equation}
	\Delta \hat{A}^2\Delta \hat{B}^2 \geq \left|\frac{1}{2} \langle [\hat{A},\hat{B}] \rangle \right|^2,
\end{equation}
where $\Delta \hat{X}$ is the variance measure of the operator $\hat{X}$, $ \langle\bullet \rangle$ is the expectation value of the  operator $\bullet $, and $ [\hat{A},\hat{B}]$ is the commutator between $\hat{A}$ and $\hat{B}$. The previous inequality is the widely known formula of uncertainty relation in quantum information processing \cite{PhysRevLett.92.117903,PhysRevLett.98.140402,RevModPhys.95.011003}.  The Schrödinger uncertainty relation is regarded as more robust as it provides tighter bounds on the product of uncertainties by including the covariance term \cite{schrodinger1935discussion}
\begin{equation} \label{schr}
		\Delta \hat{A}^2\Delta \hat{B}^2 \geq \left|\frac{1}{2} \langle [\hat{A},\hat{B}] \rangle \right|^2+ |\text{Cov}(\hat{A},\hat{B})|^2,
\end{equation}
where $\text{Cov}(\hat{A},\hat{B})=\frac{1}{2}(\hat{A}\hat{B}+ \hat{B} \hat{A})- \langle\hat{A} \rangle \langle\hat{B} \rangle$ is the covariance. \textcolor{black}{While these relations constrain the product of variances, the recent focus has shifted to the sum of variances, which can provide stronger and more versatile bounds, especially when one variance is small.} Recently, Maccone and Pati suggested a set of stronger uncertainty relations that incorporate mutually exclusive states \cite{PhysRevLett.113.260401}. Their approach allows for improved lower bounds on the sum of variances for two incompatible observables, demonstrating that these bounds are non-trivial even when one observable is an eigenstate of another. One of these \textcolor{black}{relations is} given by
\begin{equation}\label{MPU}
	\Delta \hat{A}^2+\Delta \hat{B}^2 \geq \pm i \langle [\hat{A},\hat{B}] \rangle+ \left|\langle \psi|\hat{A}\pm i \hat{B}| \psi^\bot\rangle \right|^2.
\end{equation}
Here, $|\psi\rangle $ and $|\psi^\bot\rangle $ to represent arbitrary states that are orthogonal to each other. 

On the other hand, reverse uncertainty relations have been shown to provide insights into the limitations imposed by quantum mechanics on simultaneous measurements \cite{PhysRevA.95.052117,PhysRevResearch.3.L042012}. The development of state-dependent reverse uncertainty relations has opened new avenues for research, emphasizing their importance in both theoretical and experimental contexts \cite{PhysRevResearch.2.023106,zheng2023stronger}. The reverse uncertainty relation of two incompatible observables is expressed by \cite{PhysRevA.95.052117}
\begin{equation}\label{UUR}
		\Delta \hat{A}^2+\Delta \hat{B}^2 \leq \frac{2 \Delta(\hat{A}-\hat{B})}{1-\frac{\text{Cov}(\hat{A},\hat{B})}{\Delta\hat{A}\Delta \hat{B}}} -2 \Delta\hat{A}\Delta \hat{B} \ . 
\end{equation}

\textcolor{black}{The goal of this work is to derive significantly tighter forward and reverse uncertainty relations by employing a more powerful mathematical structure that captures the underlying geometry of quantum states more completely. The standard Schrödinger relation (\ref{schr}) is derived from the Cauchy-Schwarz inequality applied to the deviation vectors $|\delta_A\rangle = (\hat{A}-\langle\hat{A}\rangle)|\psi\rangle$ and $|\delta_B\rangle = (\hat{B}-\langle\hat{B}\rangle)|\psi\rangle$. This approach is powerful but physically incomplete, as it only uses information from the inner product $\langle \delta_A|\delta_B \rangle$, which relates to the commutator and covariance. It does not, however, utilize the information contained in the norm of the sum of these vectors, $||\delta_A \pm \delta_B||$. Our work takes a different path by focusing on the triangle inequality and its sharpest known refinement, the Maligranda inequality \cite{maligranda2006simple}. The physical power of this approach stems from its ability to relate the norms of three physically distinct deviation vectors: $||\delta_A|| = \Delta A$, $||\delta_B|| = \Delta B$, and, crucially, $||\delta_A \pm \delta_B|| = \Delta(\hat{A} \pm \hat{B})$. Therefore, unlike previous methods, our framework establishes a direct, rigorous link between the uncertainties of the individual observables and the uncertainty of their combined observable. The 'angular distance' term in the Maligranda inequality is not an abstract parameter but a quantity directly determined by the interplay of these three variances, offering a complete geometric picture of the uncertainty trade-off for a given quantum state.}

In this paper, we \textcolor{black}{leverage this geometric framework to} derive four new sets of forward and reverse uncertainty relations for the sum of variances. We demonstrate their superiority over existing bounds through detailed analysis and three physical examples. Furthermore, we generalize our method to the multi-observable case using an inequality by M. Kato et al. \cite{kato2007sharp}, situating our work in the broader context of another uncertainty relations \cite{PhysRevA.89.052115, song2017stronger,baek2019entropic,chen2019tight,zheng2020multi,li2022tighter}. The paper is structured as follows: In Section 2, we derive our main results for two observables. In Section 3, we validate these relations using examples from quantum optics, SU(2), and SU(1,1) systems. Section 4 presents the generalization to multiple observables. Finally, we conclude in Section 5.

\section{Stronger Limits for Uncertainty Relations}
The angular distance between nonzero vectors $x_1$ and $x_2$ in a normed linear space was established by Maligranda as \cite{maligranda2006simple}
\begin{equation}
	\begin{split}
		\max \{\left\|x_1\right\|,\left\|x_2 \right\|\} \big(2-\mathcal{D}\big)&\geq\left\|x_1 \right\|+ \left\|x_2 \right\| - \left\|x_1 +x_2\right\| 
		\geq \min \{\left\|x_1\right\|,\left\|x_2 \right\|\} \big(2-\mathcal{D}\big)\ ,
	\end{split}
\end{equation}
where $\mathcal{D}=\left\| \frac{x_1}{\left\| x_1\right\|}+ \frac{x_2}{\left\| x_2\right\|} \right\|$ is the angular distance. 
From previous inequality, we can establish the following inequality as
\begin{equation}\label{Eq2}
	\mathcal{F}_L \geq \left\|x_1 \right\|^2+ \left\|x_2 \right\|^2 +2  \left\|x_1 \right\| \left\|x_2 \right\| \geq 	\mathcal{F}_R \ ,
\end{equation}
with 
\begin{equation}
	\begin{split}
		& \mathcal{F}_R=\bigg[\left\|x_1 +x_2\right\|+ \min\{\left\|x_1\right\|,\left\|x_2 \right\|\} \big(2-\mathcal{D}\big)\bigg]^2 \ , \\&
		\mathcal{F}_L=\bigg[\left\|x_1 +x_2\right\|+ \max\{\left\|x_1\right\|,\left\|x_2 \right\|\} \big(2-\mathcal{D}\big)\bigg]^2 \ .
	\end{split}
\end{equation} 

\textcolor{black}{
The first part of the inequality, $\mathcal{F}_L \ge ||x_1||^2 + ||x_2||^2 + 2 ||x_1|| ||x_2|| $, is directly proven by its rearrangement
\begin{equation} \label{ll}
 F_L - 2 ||x_1|| ||x_2|| \ge	||x_1||^2 + ||x_2||^2 \ .
\end{equation} 
}
\textcolor{black}{
Similarly, the right side of the inequality (\ref{Eq2}) gives 
\begin{equation}
	\left\|x_1 \right\|^2+ \left\|x_2 \right\|^2 +2  \left\|x_1 \right\| \left\|x_2 \right\| \geq 	\mathcal{F}_R\ .
\end{equation}
 However, $ \left\|x_1 \right\|^2+ \left\|x_2 \right\|^2 \geq 2  \left\|x_1 \right\| \left\|x_2 \right\|$, hence the inequality become
 \begin{equation}
 	2 (\left\|x_1 \right\|^2+ \left\|x_2 \right\|^2)\ge \left\|x_1 \right\|^2+ \left\|x_2 \right\|^2 +2  \left\|x_1 \right\| \left\|x_2 \right\| \geq 	\mathcal{F}_R \ .
 \end{equation}
 Considering the leftmost and rightmost terms directly implies the second part of the inequality
 \begin{equation} \label{ra}
 	||x_1||^2 + ||x_2||^2 \ge \frac{1}{2} F_R \ .
 \end{equation}
\\
Combining the results from Eqs. (\ref{ll}) and (\ref{ra}), we have successfully proven that the inequality (\ref{Eq2}) can be rewritten as}
 \begin{equation}\label{eq12}
 	 \mathcal{F}_L-2  \left\|x_1 \right\| \left\|x_2 \right\| \geq \left\|x_1 \right\|^2+ \left\|x_2 \right\|^2  \geq 	\frac{1}{2}\mathcal{F}_R \ ,
 \end{equation}
 
 \textcolor{black}{Hereinafter, we begin by formulating four foundational geometric inequalities stemming from the Maligranda inequality. We construct two separate classes of uncertainty principles. The first class is responsive to the covariance of the observables, while the second is determined by their commutator. Both classes are then reconstructed through an orthogonal state projection, producing a complete set of four inequalities.}
 
 \subsection{Covariance-Sensitive Relations}
 The first class of relations is derived by constructing the deviation vectors with a real relative coefficient. Let $\hat{A}$ and $\hat{B}$ be two observables in the Hilbert space of a quantum system prepared in the pure state $|\psi \rangle$. Using quantum notation, define the two vectors $x_1$ and $x_2$ as 
 \begin{equation} \label{V1}
 	|x_1\rangle= (\hat{A}- \langle \hat{A} \rangle \mathbb{I})|\psi\rangle, \quad |x_2\rangle= \pm(\hat{B}- \langle \hat{B} \rangle \mathbb{I})|\psi\rangle \ .
 \end{equation}
 
 \subsubsection{Basic Formulation}
\textcolor{black}{A direct application of Eq. (\ref{eq12}) yields our first family of relations, which uses only information from the measurement state $|\psi\rangle$.} Then, $||x_1||=\Delta \hat{A}$,  $||x_2||=\Delta \hat{B}$, $||x_1+x_2||=\Delta (\hat{A}\pm \hat{B})$, and $\mathcal{D}= \sqrt{2\pm \frac{2 \text{Cov}(\hat{A},\hat{B})}{\Delta \hat{A} \Delta \hat{B}}}$. Besides, the first limit of the uncertainty relation is
\begin{equation} \label{Un-1}
	\mathcal{U}_{1}^{\pm}(\hat{A}, \hat{B})\geq\Delta\hat{A}^2 +\Delta\hat{B}^2 \geq \mathcal{L}_{1}^{\pm}(\hat{A}, \hat{B}) \ , 
\end{equation}
where 
\begin{equation}
	\begin{split}
		&\mathcal{U}_{1}^{\pm}(\hat{A}, \hat{B}) = \left[\Delta (\hat{A} \pm \hat{B})+ G_\pm \max(\Delta\hat{A},\Delta\hat{B}) \right]^2 - 2 \Delta\hat{A} \Delta\hat{B},
		\\ &\mathcal{L}_{1}^{\pm}(\hat{A}, \hat{B}) =\frac{1}{2} \left[\Delta (\hat{A} \pm \hat{B})+ G_\pm \min(\Delta\hat{A},\Delta\hat{B}) \right]^2 \ .
	\end{split}
\end{equation}
 Here,  $G_\pm=2-\sqrt{2\pm \frac{2 \ \text{Cov}(\hat{A},\hat{B})}{\Delta \hat{A} \Delta \hat{B}}}$. This provides a complete uncertainty relation derived directly from the geometry of the state $|\psi\rangle$.
 
 \subsubsection{Formulation via an Orthogonal State $|\psi^\perp\rangle$}
  Using the vectors from Eq. (\ref{V1}) and selecting an arbitrary state vector $|\psi^\bot\rangle$ that is orthogonal to $|\psi \rangle$, the bounds can be significantly tightened by incorporating information from this orthogonal state. According to the experimental and theoretical Maccone-Pati inequality \cite{PhysRevLett.113.260401,PhysRevA.93.052108},  the optimal choice for the orthogonal state \( |\psi^\bot\rangle \) is given by $|\psi^\bot\rangle_{\hat{\mathcal{O}}} \propto( \hat{\mathcal{O}}-\langle \hat{\mathcal{O}}  \rangle  \mathbb{I} )\big|\psi\rangle$. Provided that  $|\psi\rangle$ constitutes an eigenstate of the operator $\hat{A}$ or the operator $\hat{B}$, the best orthogonal state $|\psi^\bot\rangle$ correspondingly aligns with $|\psi^\bot\rangle_{\hat{B}}$ or $|\psi^\bot\rangle_{\hat{A}}$, respectively. Conversely, if $|\psi \rangle$ does not constitute an eigenstate of either $\hat{A}$ or $\hat{B}$, and  $|\psi_\bot\rangle_{A} \neq |\psi_\bot\rangle_{B}  $, the composite operator may be formulated as $\hat{\mathcal{O}}=\hat{A}\pm\hat{B}$.  \textcolor{black}{ Using the Cauchy-Schwarz inequality with the vectors $|u\rangle = |x_1\rangle + |x_2\rangle$ and $|v\rangle = |\psi^\perp\rangle$ (where $||v||=1$), we get
 \begin{equation}
 	\Delta(\hat{A}\pm\hat{B})^2 \ge |\langle\psi^\perp | (|x_1\rangle + |x_2\rangle) \rangle|^2 = |\langle\psi^\perp|\hat{A}\pm\hat{B}|\psi\rangle|^2 \ .
 \end{equation}
 	Similarly, the angular distance term $\mathcal{D} $ can be obtained by the same way 
 	\begin{equation}
 	\mathcal{D}^2 \ge \left| \left\langle\psi^\perp \middle| \left( \frac{|x_1\rangle}{||x_1||} + \frac{|x_2\rangle}{||x_2||} \right) \right\rangle \right|^2 = \left| \left\langle\psi^\perp \middle| \left( \frac{\hat{A}}{\Delta\hat{A}} \pm \frac{\hat{B}}{\Delta\hat{B}} \right) \middle| \psi \right\rangle \right|^2 \ .
 	\end{equation}
 }
 	 Consequently, the second strengthened inequality of the uncertainty principle governing upper and lower bounds is expressed as
 \begin{equation} \label{Un-2}
 	\mathcal{U}_{2}^{\pm}(\hat{A}, \hat{B})\geq\Delta\hat{A}^2 +\Delta\hat{B}^2 \geq \mathcal{L}_{2}^{\pm}(\hat{A}, \hat{B}) \ , 
 \end{equation}
 where 
 \begin{equation}
 	\begin{split}
 		&\mathcal{U}_{2}^{\pm}(\hat{A}, \hat{B}) = \left[\mathcal{J}_\pm+ \mathcal{K}_\pm \max(\Delta\hat{A},\Delta\hat{B}) \right]^2 - 2 \Delta\hat{A} \Delta\hat{B},
 		\\ &\mathcal{L}_{2}^{\pm}(\hat{A}, \hat{B}) = \frac{1}{2}\left[\mathcal{J}_\pm+ \mathcal{K}_\pm \min(\Delta\hat{A},\Delta\hat{B}) \right]^2 \ .
 	\end{split}
 \end{equation}
Here, $\mathcal{J}_\pm=\big| \langle \psi^\bot|\hat{A}\pm \hat{B} |\psi \rangle\big|$, $\mathcal{K}_\pm=2- \bigg|\langle \psi^\bot|\frac{\hat{A}}{\Delta \hat{A}}\pm \frac{\hat{B}}{\Delta \hat{B}} |\psi \rangle \bigg|$.\\

\par 
 
 \subsection{Commutator-Sensitive Relations}
 
 \textcolor{black}{The second class of relations is designed to explicitly harness the information contained in the commutator, which quantifies the fundamental incompatibility of the observables. To explore the role of the imaginary part, which corresponds to the commutator, we introduce the definition of the vectors $x_1$ and $x_2$ as}
\begin{equation} \label{V2}
	|x_1\rangle= (\hat{A}- \langle \hat{A} \rangle \mathbb{I})|\psi\rangle, \quad |x_2\rangle= \pm i (\hat{B}- \langle \hat{B} \rangle \mathbb{I})|\psi\rangle \ .
\end{equation}

\subsubsection{Basic Formulation}
\textcolor{black}{Applying our geometric framework with this new vector definition yields
$$||x_1||=\Delta \hat{A}, \  \ ||x_2||=\Delta \hat{B},  \ ||x_1+x_2||=\Delta (\hat{A}\pm i \hat{B}),\ \  \text{and}\ \ \mathcal{D}= \sqrt{2\pm \frac{i  \langle [\hat{A}, \hat{B}] \rangle }{\Delta \hat{A} \Delta \hat{B}}}\ .$$
}
Consequently, the third set of tighter limits is given by
\begin{equation} \label{Un-3}
	\mathcal{U}_{3}^{\pm}(\hat{A}, \hat{B})\geq\Delta\hat{A}^2 +\Delta\hat{B}^2 \geq \mathcal{L}_{3}^{\pm}(\hat{A}, \hat{B}) \ , 
\end{equation}
where 
\begin{equation}
	\begin{split}
		&\mathcal{U}_{3}^{\pm}(\hat{A}, \hat{B})= \big[\mu_\pm + \nu_\pm \max\{\Delta \hat{A}, \Delta \hat{B}\} \big]^2- 2 \Delta\hat{A} \Delta\hat{B},\\&
		\mathcal{L}_{3}^{\pm}(\hat{A}, \hat{B})= \frac{1}{2} \big[\mu_\pm + \nu_\pm \min\{\Delta \hat{A}, \Delta \hat{B}\} \big]^2 \ .
	\end{split}
\end{equation}
Here, $\mathcal{\mu}_\pm=\Delta(\hat{A}\pm i \hat{B})$, and $\mathcal{\nu}_\pm=2-\sqrt{2\pm \frac{i \langle [\hat{A}, \hat{B}] \rangle}{\Delta \hat{A} \Delta \hat{B}}}$.

\subsubsection{Formulation via an Orthogonal State}
Utilizing the vectors from Eq. (\ref{V2}) and selecting eigenvector  $|\psi_\bot\rangle$ orthogonal to the state $|\psi\rangle$, the fourth inequality governing the upper and lower bounds of the uncertainty principle is given by
\begin{equation} \label{Un-4}
	\mathcal{U}_{4}^{\pm}(\hat{A}, \hat{B})\geq\Delta\hat{A}^2 +\Delta\hat{B}^2 \geq \mathcal{L}_{4}^{\pm}(\hat{A}, \hat{B})\ .
\end{equation}
Here, 
\begin{equation}
	\begin{split}
		&\mathcal{U}_{4}^{\pm}(\hat{A}, \hat{B})= \big[\Lambda_\pm + \chi_\pm \max\{\Delta \hat{A}, \Delta \hat{B}\} \big]^2- 2 \Delta\hat{A} \Delta\hat{B}\ ,\\&
		\mathcal{L}_{4}^{\pm}(\hat{A}, \hat{B})= \frac{1}{2}\big[\Lambda_\pm + \chi_\pm \min\{\Delta \hat{A}, \Delta \hat{B}\} \big]^2 \ ,
	\end{split}
\end{equation}
where $\Lambda_\pm=|\langle\psi^\bot |\hat{A}\pm i \hat{B}|\psi\rangle|,$ and $\chi_\pm=2-\left|\langle \psi^\bot|\frac{\hat{A}}{\Delta \hat{A}}\pm \frac{i\hat{B}}{\Delta \hat{B}}|\psi \rangle\right|$.

\subsection{The Combined Uncertainty Relation}

Overall, in instances where $\mathcal{L}_{i}^{+}(\hat{A}, \hat{B})\neq \mathcal{L}_{i}^{-}(\hat{A}, \hat{B})$, and $\mathcal{U}_{i}^{+}(\hat{A}, \hat{B})\neq \mathcal{U}_{i}^{-}(\hat{A}, \hat{B})$, the limits of uncertainty and reverse uncertainty relations may be combined into the form
\begin{equation} \label{Both}
	\begin{split}
			\min\{\mathcal{U}_{i}^{+}(\hat{A}, \hat{B}), \mathcal{U}_{i}^{-}(\hat{A},\hat{B})\} &\geq\Delta\hat{A}^2 +\Delta\hat{B}^2   \geq
		\max\{\mathcal{L}_{i}^{+}(\hat{A}, \hat{B}), \mathcal{L}_{i}^{-}(\hat{A},\hat{B})\}, 
	\end{split}
\end{equation}
where $ i=1,2,3,\ \text{or}\ 4$. This combination yields the strongest possible bound for the sum of variances.

\section{Examples}
This section examines the validity of existing lower bounds by employing the newly derived inequalities. Furthermore, we compare our results with those obtained from other inequalities. Three illustrative examples facilitate this comparative analysis: two operators associated with oscillating field amplitudes, two operators satisfying the SU(1,1) algebra, and two operators satisfying the SU(2) algebra.

\subsection{Oscillating Field Amplitudes}
Now, let the two operators $\hat{A}$ and $\hat{B}$ be defined based on the annihilation $(\hat{a})$ and creation $(\hat{a}^\dagger)$ as
\begin{equation}\label{Oss}
	\hat{A}=\hat{a}^\dagger \hat{a}, \quad \hat{B}= \frac{\hat{a}+\hat{a}^\dagger}{2} \ .
\end{equation}
 Physically, $\hat{A}$ is the total number of quanta in the system, and  $\hat{B}$ represents the position-like observable in quantum optics. Moreover, the superposition state measures are given by
\begin{equation}\label{psi1}
	\begin{split}
		&|\psi\rangle= \cos \theta |0\rangle+\sin \theta e^{i \phi}|1\rangle, \\
		& |\psi^\bot \rangle= \sin \theta |0\rangle -\cos \theta e^{i \phi}|1\rangle.
	\end{split}
\end{equation}
\textcolor{black}{The analytical expressions for all variances and expectation values of these two operators are provided in Appendix A.1.}

\begin{figure}[h]
	\centering
	\includegraphics[scale=0.85,trim=00 00 00 00, clip]{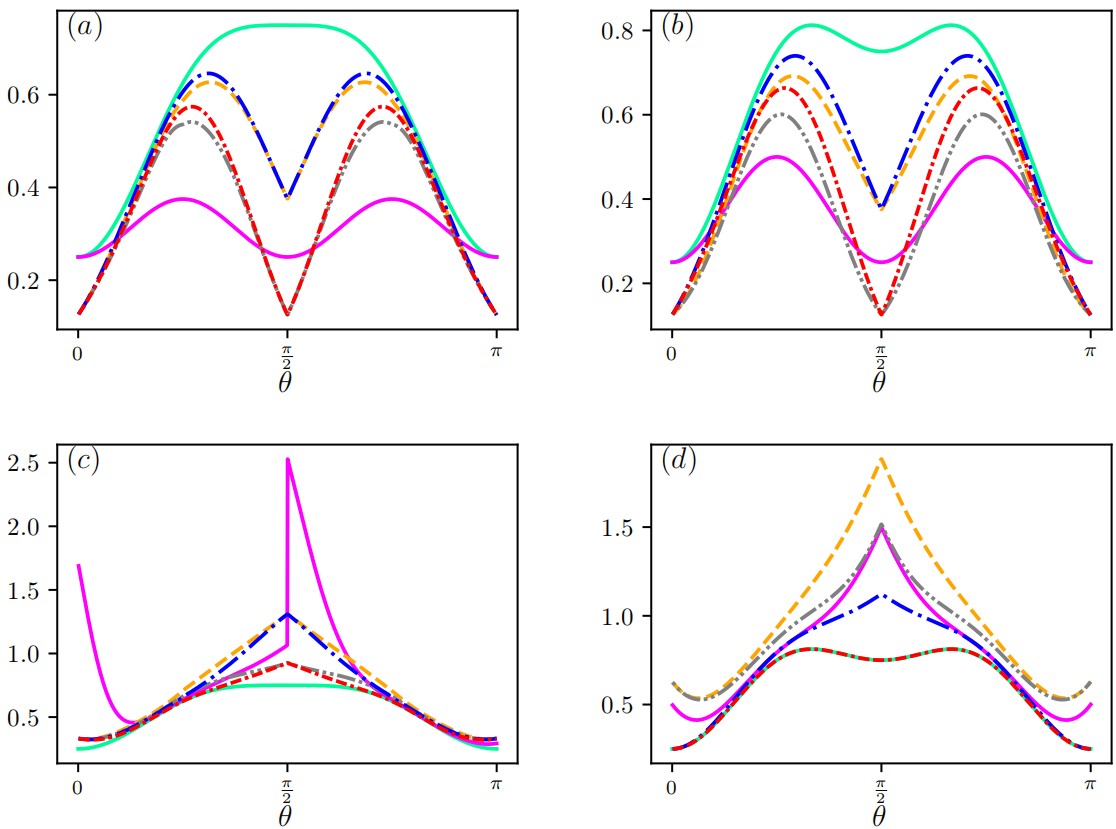} %
	\vspace{-0.1cm}
	\caption{ Uncertainty and reverse uncertainty limits for the number and quadrature operators (Eq. \ref{Oss}) as a function of the state parameter $\theta$. The green curve in all plots represents the exact sum of variances, $\Delta\hat{A}^2 + \Delta\hat{B}^2$. (a), (b) show the lower bounds for a relative phase of $\phi=\pi/4$ and $\phi=\pi/2$, respectively. The magenta curve is the Maccone-Pati bound \cite{PhysRevLett.113.260401}. The orange, grey, blue, and red curves correspond to our maximum bounds of $\mathcal{L}_1^\pm(\hat{A}, \hat{B})$, $\mathcal{L}_2^\pm(\hat{A}, \hat{B})$, $\mathcal{L}_3^\pm(\hat{A}, \hat{B})$, and $\mathcal{L}_4^\pm$, respectively. (c), (d) show the upper bounds for the same phase angles. The magenta curve is the bound from Ref. \cite{PhysRevLett.113.260401}. The orange, grey, blue, and red curves correspond to our minimum new bounds of $\mathcal{U}_1^\pm (\hat{A}, \hat{B})$, $\mathcal{U}_2^\pm(\hat{A}, \hat{B})$, $\mathcal{U}_3^\pm (\hat{A}, \hat{B})$, and $\mathcal{U}_4^\pm(\hat{A}, \hat{B})$.}
	\label{f1}
\end{figure}
The comparative behavior of the derived forward and reverse inequalities with the summation of variances governed by the observables Eq. (\ref{Oss}) is presented in Fig. \ref{f1}. \textcolor{black}{The four derived families of relations can be categorized into two distinct classes. The bounds $\mathcal{L}_{1,2}^\pm$ and $\mathcal{U}_{1,2}^\pm$ (orange and gray curves for 1 and 2) are derived from a real-vector formulation, which measures the statistical correlation of the measurement outcomes. In contrast, the bounds $\mathcal{L}_{3,4}^\pm$ and $\mathcal{U}_{3,4}^\pm$ (blue and red curves for 3 and 4) are derived using a complex phase, which quantifies the fundamental incompatibility of the observables. For the number-quadrature system, the commutator $[\hat{n}, \hat{x}] = i\hat{p}$ is a non-zero operator, signifying a robust, state-independent incompatibility.} 

For the forward uncertainty inequalities and the role of orthogonal state information, Figs. \ref{f1} (a) and (b), the proposed inequalities exhibit precise alignment with variance summation through controlled phase angle modulation.  Although the Maccone-Pati inequality remains one of the most formidable formulations in uncertainty quantification, its analytical power hinges critically on selecting an optimal orthogonal state. This state is defined as $ |\psi^\bot \rangle= \sin \theta |0\rangle -\cos \theta e^{i \phi}|1\rangle$ . In this case, the Maccone-Pati inequality demonstrates relative weakness. Even all four derived inequalities exhibit weakness at the minimum values of the variance summation. Among these, $\mathcal{L}_{3}^{\pm}(\hat{A}, \hat{B})$ (blue curve) emerges as the most robust formulation, maintaining convergence toward the left-hand side of the inequality despite phase angle variations. 

On the other hand, the reverse uncertainty relations are illustrated in FIGs. \ref{f1}(c) and (d), highlighting close agreement with the sum of the variances and the new inequalities. Modifying Eq. (\ref{UUR}) with the method from Eq.(\ref{Both}) may yield a stronger reverse uncertainty inequality. Nevertheless, a trade-off exists between $\mathcal{U}{2}^{\pm}(\hat{A}, \hat{B})$ and $\mathcal{U}{4}^{\pm}(\hat{A}, \hat{B})$, which facilitates an upper bound approximation that closely aligns with the variance sum. Although Eq. (\ref{UUR}) provides a robust upper bound at \(\phi = \pi/2\), it is important to note that \(\mathcal{U}_{4}^{\pm}(\hat{A}, \hat{B})\) surprisingly surpasses it, even matching it in some cases.  Overall, the phase parameter plays a crucial role in refining the upper bounds of reverse uncertainty, ensuring that the equivalence of both sides is maintained in at least one of the new inequalities.

\subsection{SU(2) Algebra: Spin-1 System}
\textcolor{black}{Next, we examine the performance of our relations for a system with SU(2) symmetry, a cornerstone of quantum mechanics describing angular momentum and spin. We choose the incompatible observables to be two components of the angular momentum for a spin-1 particle}
\begin{equation}
	\hat{A} = \hat{J}_x, \quad \hat{B} = \hat{J}_y.
\end{equation}
\textcolor{black}{These observables obey the well-known commutation relation $[\hat{J}_x, \hat{J}_y] = i\hbar\hat{J}_z$, indicating a fundamental, state-independent incompatibility.} We employ the states  $|\psi\rangle$ and $|\psi^\bot \rangle$, which are  expressed as follows
\begin{equation}\label{psi2}
	\begin{split}
		&|\psi \rangle= \cos \theta |-1\rangle+ \sin \theta e^{i \phi} |1\rangle, \\&
	|\psi^\bot \rangle= \frac{\sqrt{3}}{2}\big (\sin \theta |-1\rangle+ \frac{\sqrt{3}}{3} |0\rangle-\cos \theta e^{-i \phi} |1\rangle\big).
	\end{split}
\end{equation}
\textcolor{black}{The choice of a sub-optimal orthogonal state, as discussed in Ref. \cite{PhysRevLett.113.260401}, provides a stringent test for the robustness of our bounds. All necessary analytical expressions are detailed in Appendix A.2.}

\begin{figure}[h]
	\includegraphics[scale=0.805,trim=00 00 00 00, clip]{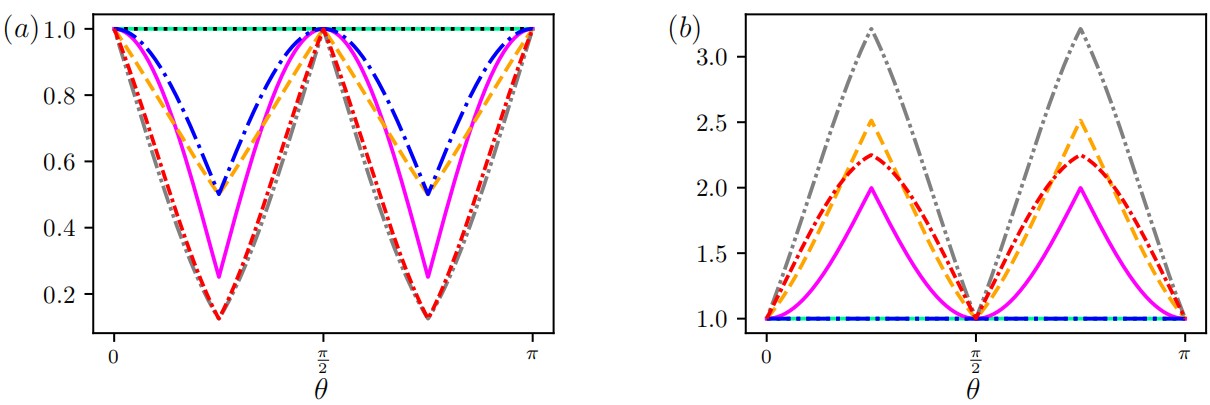} %
	\vspace{-0.1cm}
	\caption{Uncertainty limits for the angular momentum operators $\hat{J}_x$ and $\hat{J}_y$. The green curve is the exact sum of variances, $\Delta\hat{A}^2 + \Delta\hat{B}^2=1$. The black dashed curve shows that \textbf{all four} of our bounds converge and become perfectly tight at the optimal phase $\phi=\pi/2$.
		(a) shows the lower bounds for a phase of $\phi=0$. The magenta curve is the Maccone-Pati bound. The orange, grey, blue, and red curves correspond to our bounds $\mathcal{L}_1^\pm$, $\mathcal{L}_2^\pm$, $\mathcal{L}_3^\pm$, and $\mathcal{L}_4^\pm$. (b) the upper bounds for $\phi=0$. The orange, grey, blue, and red curves correspond to our bounds $\mathcal{U}_1^\pm$, $\mathcal{U}_2^\pm$, $\mathcal{U}_3^\pm$, and $\mathcal{U}_4^\pm$.
	}
	\label{f2}
\end{figure}

Figure \ref{f2} exemplifies the precision of the derived uncertainty relations within the SU(2) Lie algebraic framework, employing spin-1 systems as an illustrative case study. As evidenced, the modulation of the phase angle ($\phi$) engenders a marked amplification in the robustness of the uncertainty relation. \textcolor{black}{For a generic phase like $\phi=0$, our bounds provide non-trivial but distinct constraints. However, as the phase is tuned to $\phi=\pi/2$, a remarkable phenomenon occurs: as shown by the black dashed curve in Fig. \ref{f2}(a), \textbf{all four of our lower bounds ($\mathcal{L}_{1-4}^{\pm}(\hat{A}, \hat{B})$) converge to a single line that coincides  perfectly with the true sum of variances}. This is a central point of strength for our work. While the Maccone-Pati relation (magenta curve) remains completely insensitive to this change in phase and provides only a loose, static lower limit, our formalism correctly identifies $\phi=\pi/2$ as a special phase. Physically, this phase value aligns the geometry of the quantum state $|\psi\rangle$ in such a way that the covariance-based and commutator-based descriptions of the uncertainty become equivalent and exact. Our four distinct mathematical constructions, which probe different aspects of the state's geometry, are all unified at this point of optimal alignment, demonstrating the completeness of our geometric approach. 
	\\
	 The bounds $\mathcal{L}_3^{\pm}(\hat{A}, \hat{B})$ and $\mathcal{U}_3^{\pm}(\hat{A}, \hat{B})$ depend on $\mu_\pm = \Delta(\hat{A}\pm i\hat{B}) = \Delta(\hat{J}_x \pm i\hat{J}_y) = \Delta(\hat{J}_\pm)$. The bounds $\mathcal{L}_4^{\pm}(\hat{A}, \hat{B})$ and $\mathcal{U}_4^{\pm}(\hat{A}, \hat{B})$ depend on the projection $\Lambda_\pm = |\langle\psi^\perp|\hat{J}_\pm|\psi\rangle|$. At the specific phase $\phi=\pi/2$, the state $|\psi\rangle$ becomes an eigenstate of the operator $\hat{J}_y$. For such a state, the expectation values of the ladder operators $\langle\hat{J}_\pm\rangle$ and their squares $\langle\hat{J}_\pm^2\rangle$ take on specific values that cause the variance $\Delta(\hat{J}_\pm)$ to be precisely constrained. Simultaneously, the projection $\Lambda_\pm= |\langle \psi^\bot |J_\pm |\psi\rangle |$, which involves the orthogonal state, also becomes maximally constrained.}  Notably, in this instance, the Schr\"odinger uncertainty principle attains its tightness value at $\phi = \pi$, which coincides with the sum of the variances. Consequently, the phase angle plays a pivotal role, even under conditions characterized by weak uncertainty constraints.

\subsection{SU(1,1) Algebra}

\textcolor{black}{Our final example investigates the SU(1,1) Lie algebra, which is fundamental to describing systems involving parametric amplification, two-photon processes, and squeezed states. The generators of this non-compact group can be realized using two bosonic modes. We define our incompatible observables as two of these generators:}

\begin{equation}\label{ex3}
	\hat{A}= \hat{K}_x=\frac{\hat{a}_1^\dagger\hat{a}_2^\dagger+\hat{a}_1 \hat{a}_2}{2}, \quad \hat{B}= \hat{K}_y=\frac{\hat{a}_1^\dagger\hat{a}_2^\dagger-\hat{a}_1 \hat{a}_2}{2 i}.
\end{equation}
Using the measure states  $|\psi\rangle$ and $|\psi^\bot \rangle$ as
\begin{equation} \label{psi3}
	\begin{split}
		&|\psi \rangle= \cos \theta |00\rangle+ \sin \theta e^{i \phi} |11\rangle, \\&
		|\psi^\bot \rangle= \frac{\sqrt{3}}{2}\big (\sin \theta |00\rangle-\cos \theta e^{-i \phi} |11\rangle + \frac{\sqrt{3}}{3} |22\rangle\big).
	\end{split}
\end{equation} 
\textcolor{black}{ The necessary analytical expressions for all expectation values are provided in Appendix A.3}. 

\begin{figure}[h]
	\includegraphics[scale=0.859,trim=00 00 00 00, clip]{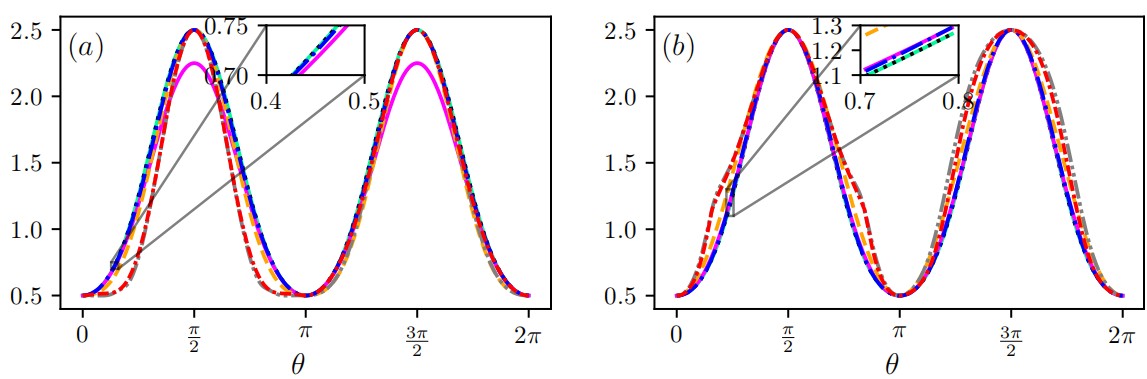} %
	\vspace{-0.5cm}
	\caption{ Uncertainty limits for the SU(1,1) operators $\hat{K}_x$ and $\hat{K}_y$. The green curve is the exact sum of variances, $\Delta\hat{A}^2 + \Delta\hat{B}^2$. The black dashed curve shows our all bounds for the optimal phase $\phi=\pi/4$, where it becomes almost perfectly tight.
		(a)  the lower bounds for a phase of $\phi=0$. The magenta curve is the Maccone-Pati bound. The orange, grey, blue, and red curves correspond to minimum values of the bounds $\mathcal{L}_1^\pm$, $\mathcal{L}_2^\pm$, $\mathcal{L}_3^\pm$, and $\mathcal{L}_4^\pm$. 
		(b) the upper bounds for $\phi=0$. The magenta curve is Maccone-Pati bound. The orange, grey, blue, and red curves correspond to the maximum values of the bounds $\mathcal{U}_1^\pm$, $\mathcal{U}_2^\pm$, $\mathcal{U}_3^\pm$, and $\mathcal{U}_4^\pm$.}
	\label{f3}
\end{figure}

The investigation into the derived relations and their robustness within the SU(1,1) quantum system—as presented in Eq. (\ref{ex3}) and Fig. \ref{f3}—indicates a notable alignment between the new forward uncertainty relations and the summation of variances. Notably, $\mathcal{L}_{1}^{\pm}(\hat{A}, \hat{B})$ and $\mathcal{L}_{3}^{\pm}(\hat{A}, \hat{B})$ display an almost exact correspondence, irrespective of the incorporation of a phase angle, thereby substantiating the efficacy of our novel uncertainty relation. Also, $\mathcal{L}_{2}^{\pm}(\hat{A}, \hat{B})$ and $\mathcal{L}_{4}^{\pm}(\hat{A}, \hat{B})$ exhibit occasional deviations from the sum of the variances; however, they ultimately converge towards it, achieving a precise concordance. Conversely, the reverse uncertainty relations, $\mathcal{U}_{3}^{\pm}(\hat{A}, \hat{B})$, exhibit a near-perfect alignment with the summation of variances. The inset plot indicates that this relation provides the tightest bound for the variance sum of the operators $\hat{A}$ and $\hat{B}$. Furthermore, it exhibits superior performance when compared to the constraints specified in Eq. (\ref{UUR}), suggesting that the bounds provided by $\mathcal{U}_{3}^{\pm}(\hat{A}, \hat{B})$ are the tighter limit. Additionally, the incorporation of the phase angle universally enhances the precision of the uncertainty bounds across all formulations. Controlling the phase angle refines the accuracy of the uncertainty limits, making the reverse uncertainty relations more robust and reliable.

\textcolor{black}{ This example underscores the importance of having a diverse toolkit of uncertainty relations. While the "un-strengthened" commutator bound $\mathcal{L}_3^{\pm}(\hat{A}, \hat{B})$ was optimal for the SU(2) case, the bounds $\mathcal{L}_1^{\pm}(\hat{A}, \hat{B})$ and  $\mathcal{L}_3^{\pm}(\hat{A}, \hat{B})$  are superior here. This shows that no single formulation is universally optimal for all physical systems. The strength of our work lies in providing a comprehensive set of four distinct, geometrically motivated relations. By comparing them, one can identify the tightest possible bound for a given physical system, whether the dominant physics is governed by simple incompatibility (as in SU(2)) or by more complex, multi-mode correlations (as in SU(1,1)).}

\section{Uncertainty Relation for Multi-Observable}\label{sec2}
If $\{x_l\}_{l=1,2,3,...,n}$ are $n$-nonzero vectors in a normed linear space, M. Kato et al. refinement of the generalized triangle inequality as \cite{kato2007sharp}
\begin{equation}\label{KKa}
	\begin{split}
		\max_{k\in \{1,\dots n\}}\{\left\|x_k \right\|\} \left(n-\left\| \sum_{j=1}^{n} \frac{x_j}{\left\| x_j\right\|} \right\|\right)\geq\sum_{j=1}^{n} \left\|x_j \right\| - \left\|\sum_{j=1}^{n}x_j \right\| \geq \min_{k\in \{1,\dots n\}}\{\left\|x_k \right\|\} \left(n-\left\| \sum_{j=1}^{n} \frac{x_j}{\left\| x_j\right\|} \right\|\right).
	\end{split}
\end{equation}

Via the Cauchy–Schwarz inequality $$\bigg(\sum_{j=1}^{n} \left\|x_j \right\| \bigg)^2 \leq n \sum_{j=1}^{n} \left\|x_j \right\|^2.$$
Via squared Eq. (\ref{KKa}) and using Cauchy–Schwarz inequality, then

	\begin{equation} \label{general}
		\begin{split}
			\Bigg[\left\|\sum_{j=1}^{n}x_j \right\|  +  \Xi \max_{k}\{\left\|x_k \right\|\}\Bigg]^2- \sum_{i\neq j}^{n} \left\|x_i \right\| \left\|x_j \right\|	\geq\sum_{j=1}^{n} \left\|x_j \right\|^2 \geq \frac{1}{n}& \Bigg[\left\|\sum_{j=1}^{n}x_j \right\|  +  \Xi \min_{k}\{\left\|x_k \right\|\}\Bigg]^2 \ ,
		\end{split}
	\end{equation}
where, $\Xi=\left(n-\left\| \sum_{j=1}^{n} \frac{x_j}{\left\| x_j\right\|} \right\|\right)$.

\textcolor{black}{The Kato inequality provides bounds on the norm of a sum of $n$ vectors $\{x_j\}_{j=1}^n$. Following the same logic as in Section 2, this can be translated into an inequality for the sum of squared norms. For a set of $n$ observables $\{\hat{A}_j\}$, we define a corresponding set of deviation vectors }
\begin{equation} \label{sg}
	|x_j\rangle= a_j \bar{A}_j |\zeta \rangle,
\end{equation} 
where $a_j$ equal $\pm 1$ or $\pm i$, and $\bar{A}_j= \hat{A}_j- \langle \hat{A}_j \rangle \mathbb{I}$. \textcolor{black}{ These constants define the specific linear combination of observables whose uncertainty is being bounded. Physically, they represent the experimenter's freedom to choose the relative signs and phases to find the tightest possible bound for a given set of observables.}
 Consequently, the multi-observable uncertainty relation limits can be obtained by
\begin{equation} \label{Gen1}
	\begin{split}
		\mathbb{U} \geq\sum_{j=1}^{n} \Delta \hat{A}_j^2 \geq \mathbb{L},
	\end{split}
\end{equation}
\begin{equation*}
	\begin{split}
	&\mathbb{U}=\Bigg[\mathbb{F}  +  \left(n- \mathbb{G} \right)\max_{k}\{\Delta \hat{A}_k\}\Bigg]^2- \sum_{i\neq j}^{n} \Delta \hat{A}_i \Delta \hat{A}_j \\&
	\mathbb{L}= \frac{1}{n} \Bigg[\mathbb{F} +  \left(n- \mathbb{G} \right)\min_{k}\{\Delta \hat{A}_k\}\Bigg]^2,
	\end{split}
\end{equation*}
with $\mathbb{F}= \Delta \big( \sum_{j=1}^{n} a_j \hat{A}_j \big)$, and $\mathbb{G}= \sqrt{n+\sum_{i\neq j}^{n}\frac{a^*_i a_j \langle \bar{A}_i \bar{A}_j\rangle}{\Delta \hat{A}_i \Delta \hat{A}_j}}$.
 
On the other hand, if we defined the state $|\zeta^\bot\rangle$ to be orthogonal on the state $|\zeta\rangle$, the summation of multi-observable uncertainty can be expressed as 
\begin{equation} \label{Gen2}
	\begin{split}
		\mathbb{U}^\bot \geq\sum_{j=1}^{n} \Delta \hat{A}_j^2 \geq \mathbb{L}^\bot,
	\end{split}
\end{equation}
\begin{equation*}
	\begin{split}
&\mathbb{U}^\bot=\Bigg[\mathbb{F}^\bot   +  \left(n- \mathbb{G}^\bot
	\right)\min_{k}\{\Delta \hat{A}_k\}\Bigg]^2- \sum_{i\neq j}^{n} \Delta \hat{A}_i \Delta \hat{A}_j\\&
	\mathbb{L}^\bot= \frac{1}{n} \Bigg[\mathbb{F}^\bot   +  \left(n- \mathbb{G}^\bot
		\right)\min_{k}\{\Delta \hat{A}_k\}\Bigg]^2.
	\end{split}
\end{equation*}
Here, $\mathbb{F}^\bot=\bigg|\langle\zeta^\bot | \sum_{j=1}^{n} a_j \hat{A}_j|\zeta\rangle \bigg|$, and $\mathbb{G}^\bot= \bigg|\bigg\langle \zeta^\bot\bigg|\sum_{j}^{n}\frac{a_j\hat{A}_j}{\Delta \hat{A}_j}\bigg|\zeta \bigg\rangle\bigg|$.

In fact, Eqs (\ref{Gen1}) and (\ref{Gen2}) constitute generalized formulations of the uncertainty principle for multiple observables, wherein the constants $a_j$ afford complete freedom in determining the strength of the uncertainty relation. However, an additional generalization exists, which involves rendering these constants dependent upon the angle between the observables. This generalization could potentially enhance the strength of the uncertainty relation. However, it necessitates further investigation, an aspect that has not been explored within the scope of this study.

\section{Conclusion}\label{conclusion}

In this study, we have introduced a novel geometric framework, founded on the Maligranda inequality, to derive four new families of uncertainty and reverse uncertainty relations for the sum of variances. \textcolor{black}{By organizing these relations into two distinct physical classes—one sensitive to the covariance and the other to the commutator—we have created a comprehensive toolkit for quantifying quantum uncertainty.}

Our investigation across three canonical physical systems (oscillating fields, SU(2), and SU(1,1)) yielded significant insights. We demonstrated that the optimal type of uncertainty relation is system-dependent: for systems like SU(2) with a strong, fundamental commutation relation, our commutator-sensitive bounds proved far superior. For systems with complex multi-mode correlations like SU(1,1), the tightest bounds were those independent of projections onto an orthogonal state. \textcolor{black}{The most powerful result of this work is the demonstration of perfect bound saturation. We showed that for specific, geometrically optimal quantum states, our relations become equalities, providing an exact prediction for the sum of variances. This feature, which is not present in other prominent sum-based relations, serves as an analytical proof of our method's precision.}

Furthermore, by extending our methodology and using the inequality form of M. Kato et al., we established two generalized forms of the sum forward and reverse uncertainty relations for multi-observable settings.

\newpage
\appendix
\section{Appendix A: Analytical Expressions for Examples}
\textcolor{black}{This appendix provides the explicit analytical forms for the variances, covariances, and other relevant expectation values used to generate the figures in Section 3. These expressions are calculated for the specific states and observables defined in each example.}

\subsection{Example 1: Oscillating Field Amplitudes}

\begin{align*}
	&\Delta\hat{A}^2 = \cos^2\theta \sin^2\theta, \qquad \qquad\qquad \
	\Delta\hat{B}^2 = \frac{1}{4}\left( \sin^2\theta(3-4\cos^2\theta\cos^2\phi) + \cos^2\theta \right) \\
	&\text{Cov}(\hat{A}, \hat{B}) = \frac{1}{8}\sin(4\theta)\cos\phi, \quad \qquad
	\Delta\hat{A}^2 + \Delta\hat{B}^2 = \frac{1}{4}\left( \cos^2\theta + 3\sin^2\theta + \sin^2(2\theta)\sin^2\phi \right) \\
	&\langle[\hat{A}, \hat{B}]\rangle = -i\cos\theta\sin\theta\sin\phi \qquad \qquad 
	\langle\psi^\perp|\hat{A}|\psi\rangle = -\cos\theta\sin\theta \\
	&\langle\psi^\perp|\hat{B}|\psi\rangle = \frac{-1}{2}\left( \cos(2\theta)\cos\phi + i\sin\phi \right)
\end{align*}

\subsection{Example 2: SU(2) Algebra}

\begin{align*}
	&\Delta\hat{A}^2 = \frac{1}{2}(1 - \sin(2\theta)\cos\phi), \qquad \qquad \
	\Delta\hat{B}^2 = \frac{1}{2}(1 + \sin(2\theta)\cos\phi), \\
	&\text{Cov}(\hat{A}, \hat{B}) = \sin\theta\cos\theta\sin\phi, \qquad \qquad \
	\Delta\hat{A}^2 + \Delta\hat{B}^2 = 1, \\
	&\langle[\hat{A}, \hat{B}]\rangle = i\langle\hat{J}_z\rangle =-i\cos(2\theta), \qquad \qquad \
	\langle\psi^\perp|\hat{A}|\psi\rangle = \frac{1}{2\sqrt{2}}\left( \cos\theta + e^{-i\phi}\sin\theta \right), \\
	&\langle\psi^\perp|\hat{B}|\psi\rangle = \frac{-i}{2\sqrt{2}}\left( \cos\theta - e^{-i\phi}\sin\theta \right)\ .
\end{align*}

\subsection{Example 3: SU(1,1) Algebra}

\begin{align*}
	&\Delta\hat{A}^2 = \frac{1}{16}\left( 11 - 2\sin^2(2\theta)\cos(2\phi) - 8\cos(2\theta) + \cos(4\theta) \right), \\
	&\Delta\hat{B}^2 = \sin^2\theta(1-\cos^2\theta\sin^2\phi) + \frac{1}{4}, \\
	&\text{Cov}(\hat{A}, \hat{B}) = -\cos^2\theta\sin^2\theta\sin\phi\cos\phi \\
	&\Delta\hat{A}^2 + \Delta\hat{B}^2 = \frac{1}{8}\left( 11 - 8\cos(2\theta) + \cos(4\theta) \right), \\
	&\langle[\hat{A}, \hat{B}]\rangle = i\langle\hat{K}_z\rangle = \frac{i}{2}(\cos(2\theta)-2), \\
	&\langle\psi^\perp|\hat{A}|\psi\rangle = \frac{1}{4}e^{-i\phi}\left( \sin\theta(\sqrt{3}\sin\theta+2) - \sqrt{3}e^{2i\phi}\cos^2\theta \right), \\
	&\langle\psi^\perp|\hat{B}|\psi\rangle = \frac{-i}{4}e^{-i\phi}\left( \sin\theta(\sqrt{3}\sin\theta-2) + \sqrt{3}e^{2i\phi}\cos^2\theta \right).
\end{align*} 

\section*{acknowledgments}
	This work was supported in part  by the National Natural Science Foundation of China(NSFC) under the Grants 12475087 and 12235008; by the University of Chinese Academy of Sciences.

\bibliographystyle{unsrt}
\bibliography{bm}

\end{document}